\documentclass{ws-ijmpe}
\usepackage[super,compress]{cite}
\usepackage{graphicx}
\begin{document}

\markboth{G. J. Mathews et al.}{Introduction to BBN}

%%%%%%%%%%%%%%%%%%%%% Publisher's Area please ignore %%%%%%%%%%%%%%
\catchline{}{}{}{}{}
%%%%%%%%%%%%%%%%%%%%%%%%%%%%%%%%%%%%%%%%%%%%%%%%%%%%%%%%%%%%%%%%%%%

\title{INTRODUCTION TO BIG BANG NUCLEOSYNTHESIS AND MODERN COSMOLOGY
}

\author{\footnotesize Grant J. Mathews, Motohiko Kusakabe}

\address{Department or Physics, University of Notre Dame, Center for Astrophysics\\
Notre Dame, IN  46556,
USA\\
gmathews@nd.edu, mkusakab@nd.edu}

\author{Toshitaka Kajino}

\address{Division of Theoretical Astronomy, National Astronomical Observatory
of Japan, Mitaka, Tokyo 181-8588, Japan \\
Department of Astronomy, Graduate School of Science, University of
Tokyo,  Hongo, Bunkyo-ku, Tokyo 113-0033, Japan\\
kajino@nao.ac.jp
}
\maketitle

\begin{history}
\received{1 May 2016}
\revised{6 June 2017}
%\accepted{Day Month Year}
%\comby{(xxxxxxxxxx)}
\end{history}

\begin{abstract}
Primordial nucleosynthesis remains as one of the pillars of modern cosmology. It is the testing ground 
upon which many cosmological models must ultimately rest. It is our only probe of the
universe during the important 
radiation-dominated epoch in the first few minutes of cosmic expansion.  This chapter reviews the basic equations
of space-time, cosmology, and big bang nucleosynthesis. We also summarize the current state
of observational constraints on primordial abundances along with the key nuclear reactions and
their uncertainties. We  summarize which nuclear measurements are most crucial during
the big bang. We also review various cosmological models and their
constraints. In particular, we analyze the constraints that big bang nucleosynthesis
places upon the possible time variation of fundamental constants, along with constraints on
the nature and origin of dark matter and dark energy, long-lived supersymmetric particles,
 gravity waves, and the primordial magnetic field.
\end{abstract}

\ccode{26.35.+c, 95.30.Cq, 95.35.+d, 98.65.Dx, 98.70.Vc, 98.80.Bp, 98.80.Cq, 98.80.Es, 98.90.Ft,98.80.Qc}

\section{Introduction}

%\ifthenelse{\equal\selectedlayoutstyle{6x9}}{\par\bfseries 

 We are at a unique period in the history of the human understanding of the cosmos. For the first time, we
have a clear picture of what the universe is comprised of, how long it has been in existence, and how it will evolve
in the future. This knowledge is the culmination of investigations via a number of cosmological probes including
supernovae, observations of the large scale distribution of galaxies and the inter-galactic medium, analysis of  the
cosmic microwave background, observations of the first stars of the early universe, and studies of the nucleosynthesis of the 
elements in the first few moments of cosmic expansion.  In this chapter, we review a number of outstanding questions in contemporary cosmology 
 and highlight  the input that big bang nucleosynthesis (BBN) provides toward answering them.
 
\subsection{What is the universe made of?}
The best current understanding of the composition of the universe comes from the most recent analysis from the {\it Planck Surveyor} \cite{PlanckXIII}.  If we denote the fraction of the mass
density of the universe in any form of matter, $i$, as 
$\Omega_i$, then the  temperature and polarization data alone imply \cite{PlanckXIII}  that that the fraction of the mass-energy of the universe in the normal
baryonic matter with which we are familiar is $\Omega_b h^2 = 0.02225 \pm 0.00016$,  where $h = 0.6727 \pm 0.0066$ is the present hubble parameter in units of 100 km s$^{-1}$ Mpc$^{-1}$.  However, if one adopts the combined analysis\cite{PlanckXIII} of the Planck temperature plus polarization and gravitational lensing data \cite{lensing1,lensing2}, plus  Baryon Acoustic Oscillations  (BAO) \cite{BAO} in the matter power spectrum, and supernova data from the Joint Light-curve Analysis (JLA) \cite{JLA} of Type Ia supernovae, then one obtains  $\Omega_b h^2 = 0.02230 \pm 0.00014$,  where $h = 0.6774 \pm 0.0066$.  That is the value adopted for this review.

\subsection{Baryon to photon ratio}
A key quantity for BBN is the present ratio $\eta$ of the baryon number density to the photon number density.
This quantity relates to the value of $\Omega_b h^2$ deduced by {\it Planck}.  
Specifically,
\begin{equation}
\eta \equiv \frac{n_b}{n_\gamma} \eta \approx 2.738 \times 10^{-8} \Omega_b h^2 = 6.11 \pm 0.04 \times 10^{-10}~~.
\label{etaeq}
\end{equation}
To see where this comes from, consider that  the baryon number density is
\begin{equation}
n_b = \frac{\rho_b}{m_b c^2}
\label{nbeq}
\end{equation}
and the photon number density is
\begin{equation}
n_\gamma = \frac{2 \zeta(3)}{\pi^2 \hbar^3} \biggl(\frac{kT}{c}\biggr)^3 = 410.7 \biggl(\frac{kT}{2.7255}\biggr)^2
\label{gammadens}
\end{equation}
where, a present day CMB background temperature of $T= 2.7255 (6)$ K is adopted from Ref.~[\refcite{Fixen09}].

Now, the baryon density relates to the critical density defined below.  $\rho_b = \Omega_b \rho_{\rm crit}$, with
\begin{equation}
\rho_{\rm crit} = \frac{2 H_0^2}{8 \pi G} = 1.05375(13) \times 10^{-2}h^2 ~{\rm MeV} c^{-2}{\rm cm}^{-3}~~.
\end{equation}
A subtle point about Eq. (\ref{nbeq}), however,  is that the mean baryon mass $m_b$ depends upon the composition \cite{Steigman06}.
Assuming a composition of nearly pure  hydrogen and helium at the end of  BBN one can write:
\begin{eqnarray}
m_b &=& m_H \biggl[ 1 - \biggl(1-\frac{1}{4} \biggl(\frac{m_{He}}{m_H} \biggr)\biggr)Y_p\biggr] \nonumber \\
&=& 937.11 - 6.683(Y_p - 0.25)  ~~{\rm MeV}/c^2
\end{eqnarray}
Combining these equations  leads to the baryon to photon
\begin{equation}
\eta = 2.733 \Omega_b h^2( 1 + .00717 Y_p) \biggl(\frac{2.7255}{kT}\biggr)^3
\end{equation}
If one adopts $Y_p = 0.25$ the above numerical value of Eq.~(\ref{etaeq}) results.
From Eq.~(\ref{gammadens}) one must conclude that there about 400 photons per cm$^3$ in the universe and there are roughly 2 billion photons per baryon.
This number was fixed in the epoch of baryogenesis.\cite{Sakarov91}  However, exactly how this number arises is not yet fully understood \cite{baryogenesis}.

\subsection{Dark Matter, Dark Energy, Relativistic Particles}

Based upon the {\it Planck} analysis \cite{PlanckXIII} then one can deduce that the contribution to closure from baryonic matter is  $\Omega_b  = 0.0486 \pm 0.011$.  
The total matter content, however is $\Omega_m = 0.3089 \pm 0.0062$.  This
implies that a much larger fraction of the
universe is made of a completely unknown component of "cold dark matter," 
$\Omega_c = 0.260\pm 0.006$. Even more surprising
is that the universe is predominantly made of a completely exotic form of mass energy, i.e.~dark energy, denoted  as, 
$\Omega_\Lambda = 0.691 \pm 0.006$ for the case of a cosmological constant. 
In addition to these,  there is presently an almost negligible mass-energy contribution from relativistic
photons and neutrinos which we designate as 
$\Omega_\gamma  = \rho_\gamma/\rho_c =  5.46(19) \times 10^{-5.}$ 

The present composition of the universe  is a great cosmic mystery. We are faced with the
dilemma that more than 95\% of the universe involves forms of matter  of which we know very little  about. 
It is a challenge, therefore,
to the next generation of astrophysicists and cosmologists to unravel this cosmic puzzle. Below we will explore some of the insight that BBN can place on dark energy and dark matter.

In order to attack this, however, we must first summarize why one
believe there is dark matter and dark energy. To do this we must first review the description for the dynamics of space
and time.

The precision with which these
parameters are now known is much better than a decade ago, however, one must keep in mind that that there remain differences between the {\it WMAP} analysis \cite{WMAP9} and the latest results from the Planck Surveyor \cite{PlanckXIII}.  
The reader is referred to the contributions by A. Coc and E Vangioni \cite{Coc17} and R. Nakamura et al. \cite{Nakamura17} in this volume as well as  the review in Ref.~[\refcite{Cyburt16}] for perspectives on the impact of the {\it Planck} data.
 Even so, one should also keep in mind that  these parameters are based upon the simplest
possible $\Lambda$CDM cosmology, and the analysis must be redone for more complicated cosmologies such as those described
below and in subsequent chapters. Also, besides the discrepancies between {\it WMAP} and {\it Planck}, there are still some discrepancies between the best fit cosmology and the observations. For example there is evidence of a suppression of the lowest multipole moments of the CMB. This may suggest \cite{Mathews15a,Mathews15b} 
evidence for super-horizon pre-big bang curvature possibly relating to our connection to the greater multiverse \cite{Mersini-Houghton09}.

\subsection{Elementary Dynamics of Space-time}
The great mystery of Einstein's {\it special} theory of relativity is that space = time. That is, time must be treated as a
spatial dimension, and the only invariant interval is a 4-dimensional distance including both space and time, i.e. in flat Minkowski space-time the
differential distance between two events is given by:
\begin{equation}
ds^2 = -dt^2 + dx^2 + dy^2 + dz^2,
\end{equation}
where we now use geometric units so that time is in geometrical units, $ t \rightarrow ct$. 

This is
mysterious enough, but an even greater mystery is given by the Einstein {\it general} theory of relativity which states that:
Geometry = Mass \& energy.
In tensor language this statement becomes:
$
G_{\mu \nu} = 8 \pi G T_{\mu \nu},
$
where $G_{\mu \nu} $ is the Einstein tensor describing the essential curvature of space time, while $T_{\mu \nu}$ is a tensor describing the
distribution and flow of mass and energy density through space-time. This equation does not make sense. As is often the case in modern physics.  Nevertheless, it describes reality.

If one adopts the notion that the universe is homogeneous and isotropic on a large enough scale, then space-time intervals are given by the Lemaitre-Friedmann-Robertson-Walker metric,
\begin{equation}
ds^2 = -dt^2 + a(t)^2\biggl[\frac{dr^2}{1 - kr^2} + r^2d\theta^2 + r^2 \sin^2{\theta} d\phi^2\biggr]~~,
\end{equation}
where now $a(t)$ is the dimensionless scale factor and $k$ is the curvature parameter.  For this metric,  the evolution
of the early universe is simply given by the Friedmann equation which describes the the Hubble parameter H in terms
of densities $\rho$, curvature $k$, the cosmological constant $\Lambda$, and the cosmic scale factor $a$:
\begin{equation}
\biggl(\frac{\dot a}{a}\biggr)^2 \equiv H^2 = \frac{8}{3} \pi G \rho -\frac {k}{a^2} + \frac{\Lambda}{3}
 =  H_0^2\biggl[ \frac{\Omega_\gamma}{a^4}
 +\frac{\Omega_m}{a^3}
+ \frac{\Omega_k}{a^2}
+ \Omega_\Lambda \biggr]~~,
\label{Friedmann}
\end{equation}
where $H_0 = 67.74 \pm 0.46$ km sec$^{-1}$ Mpc$^{-1}$is the present value \cite{PlanckXIII}  of the Hubble parameter.  One can then define the various closure contributions
from relativistic particles, nonrelativistic matter, curvature, and dark energy:
\begin{equation}
\Omega_\gamma = 8 \pi G \rho_\gamma/(3 H_0^2)~~, 
\end{equation}
\begin{equation}
 \Omega_m = 8 \pi G \rho_m/(3 H_0^2)~~,  
 \end{equation}
 \begin{equation} 
\Omega_k = k/(a^2 H_0^2)~,
\end{equation}
  and 
  \begin{equation}
  \Omega_\Lambda = \Lambda/(3 H_0^2)~~.
  \end{equation}

From Eq.~(\ref{Friedmann})  it is clear that as we
look into the past when the scale factor was smaller, the relative contributions of various components would have
been different. Indeed, if we go back to a time when the scale factor was about half of its present value, then matter and dark
energy contributed equally. If we move back to when the universe was only 1/1100 of its present size, then the universe is ionized into photons, electrons, and baryons.  This is the surface of last scattering when the CMB photons were emitted.  Moving further back to a time when the universe was only 1/3400 of its present size, then relativistic
photons and neutrinos dominate the mass energy of the universe. Indeed,  this is the radiation dominated epoch during  which
the universe expanded at a a  rate $a(t) \propto t^{-1/2}$.  One might want to  call this epoch the "big bang", although that term by convention
 refers to the mathematical
model describing the overall expansion of space time throughout the distant past and into the future.

The early universe includes the Planck epoch, the birth of space-time, inflation, reheating, a variety of cosmic
phase transitions (e.g. supersymmetry breaking, baryogenesis, the electroweak transition, and the QCD transition),
the epoch of big bang nucleosynthesis (BBN), and the production of the cosmic microwave background (CMB).
For most of the big bang only the radiation [$\Omega_\gamma$ term in Eq.~(\ref{Friedmann})] is important. There are, however, interesting
variants of big bang cosmology where this is not the case. 

The only direct probe of the radiation dominated epoch
is the yield of light elements from BBN in the temperature regime from $10^8$ to $10^{10}$ K and times of about 1 to $10^3$
sec into the big bang. The only other probe is the spectrum of temperature 
fluctuations in the CMB which contains information of the first quantum 
fluctuations in the universe, and the details of the distribution and evolution of dark matter, baryonic
matter, photons and electrons near the time of the surface of photon last scattering (about $2.8 \times 10^5$ yr into the big bang).

\subsection{Epoch of BBN}
One of the most powerful aspects of BBN is the simplicity\cite{Wagoner73,Yang84,Malaney93,Iocco09,Cyburt16} of the equations.  For the most part one can assume thermodynamic equilibrium of all species present.  The only non-equiulibrium processes are the thermonuclear reactions themselves.  
For all times relevant to standard  big-bang nucleosynthesis, one can ignore curvature and dark energy. Hence, the Friedmann equation is just:
\begin{equation}
\biggl(\frac{\dot a}{a}\biggr)=\sqrt{ \frac{8}{3} \pi G \rho_\gamma}  = H_0 \frac{\sqrt{\Omega_\gamma}}{a^2} \approx 1.13 T_{\rm MeV}^2 ~ {\rm sec}^{-1}  ~~.
\end{equation}
This leads to a simple relation for the evolution of the scale factor with time,
\begin{equation}
a = \Omega_\gamma^{1/4} (2 H_0 t)^{1/2}
\end{equation}

Also, at the time of BBN the timescale for Compton scattering is short and most background species are in  thermal equilibrium.  Hence, the number density of particles of type $i$ with momenta
between $p$ and $p + dp$ is simply given by Fermi-Dirac or Bose-Einstein distributions,
\begin{equation}
n_i (p)dp =  = \frac{g_i}{2 \pi^2}p^2 \biggl[ \exp\biggl(\frac{E_i(p) - \mu_i}{kT} \pm 1 \biggr]^{-1} dp
\label{ndens}
\end{equation}
where $E_i(p)$ is the energy of the particle, the $\pm$  sign is negative for bosons and positive  for fermions, while $g_i$ 
is the number of degenerate spin states of the particle (e.g.~$g = 1$ for neutral massless leptons, and $g_i = 2$ for charged leptons and photons). 

Integrating the product $E_i(p)n_i(p)$ over all momenta gives the energy density for particle $i$.
 In the standard big bang, with all chemical potentials $\mu_i$  set to zero, the total mass-energy density of the universe at the epoch of nucleosynthesis is given by
\begin{equation}
\rho = \rho_\gamma + \rho_{\nu_i} + \rho_i ~~,
\end{equation}
where $\rho_\gamma$, $\rho_{\nu_i}$, and $\rho_i$ 
are the energy densities due to photons, neutrinos, and charged leptons, respectively (including antiparticles).  [Note that during BBN, the baryons contribute negligibly to the total mass energy density.]
At the temperatures of primordial nucleosynthesis only electrons (and positrons) contribute to the charged leptons. The contribution from neutrinos is given by
(assuming only relativistic particles)
\begin{equation}
\rho_{\nu_i} = \int p[n_{\nu_i}(p) + n_{\bar \nu_i}(p)] dp~~ = \frac{7}{8} \frac{\pi^2}{15} (T_{\nu_i})^4
\end{equation}
where $T_{\nu_i}$ is the neutrino temperature and here and for the rest of the manuscript we adopt natural units, i.e. $k = c = \hbar = 1$. The energy densities from photons $\rho_\gamma$ and relativistic charged
leptons $\rho_i$ are given by
\begin{equation}
\rho_\gamma = \frac{4}{7} \rho_i = \frac{\pi^2}{15}(T_\gamma)^4 ~~,
\end{equation}
where  $T_\gamma$ is the photon temperature (henceforth we absorb the Boltzmann constant k into the temperature). For temperatures above 1 MeV,
 the neutrinos and photons are in thermal equilibrium ($T_\gamma = T_{\nu_i} = T$). 
 
 The Friedmann equation  plus the fact that temperature decreases inversely with the scale factor, can be used to derive the relation between temperature and time during BBN.
\begin{equation}
 T \approx 1.4 g_{\rm eff}^{-1/4} \biggl( \frac{t}{\rm 1~sec}\biggr)^{-1/2}~{\rm MeV}~,
 \end{equation}
 where $g_{\rm eff}(T)$ is the effective number of relativistic degrees of freedom in bosons and fermions,
 \begin{equation}
 g_{\rm eff}(T) = \sum_{\rm Bose} g_{\rm Bose} + \frac{7}{8}\sum_{\rm Fermi} g_{\rm Fermi}  ~~.
\label{geff} 
\end{equation}

 This completes the equations of the basic thermodynamics.  The nuclear reactions, however, must be followed in detail as they fall out of equilibrium.  
 for nuclide $i$ undergoing reactions of the type 
$i + j \leftrightarrow j + k$ one can write:
 \begin{equation}
 \frac{dY_i}{dt} = \sum_{i,j,k} N_i \biggr( \frac{Y_l^{N_l} Y_k^{N_k}}{N_l ! N_k ! } n_k \langle \sigma_{l k} v \rangle - 
  \frac{Y_i^{N_i} Y_j^{N_j}}{N_i ! N_j ! } n_j \langle \sigma_{i j} v \rangle\biggr)
\label{nucrates}
 \end{equation}
where $Y_i = X_i/A_i$ is the abundance fraction,  $N_i$ is the number of reacting identical particles, $n_i$ is the number density of nucleus $i$ and $\langle  \sigma_{i j} v \rangle $
denoted the maxwellian averaged reaction cross section,
\begin{equation}
\langle\sigma_{i j} v\rangle=\sqrt{\frac{8}{\pi\mu}}\left(T\right)^{-\frac{3}{2}}\int_{0}^{\infty}\sigma_{i j}\left(E\right)\: e^{-E/T}\: E\: dE~~,
\label{eq:Forward Reaction Rate}
\end{equation}
where $\mu_{i j} = m_i m_j/(m_i + m_j)$ is the reduced mass of the reacting system.
Once the forward reaction rate is known, the reverse reaction rate can be deduced from an application (cf. Ref.~[\refcite{Mathews11}]) of the principle of detailed balance. For a two body reaction one can write: 
\begin{equation}
\langle\sigma v\rangle_{l k}=\frac{g_{i}g_{j}}{g_{l} g_k (1 + \delta_{12})} \biggl(\frac{\mu_{i j}}{\mu_{k l}}\biggr)^{3/2}\langle \sigma_{12} v \rangle \exp{\bigl[ -Q/T \bigl]}~~,
\label{eq:Inverse Reaction Rate 2}
\end{equation}
where $Q$ is the energy released in the reaction.
The reaction rates relevant to BBN have been  conveniently tabularized in analytic form in several sources \cite{Cyburt10,CF88,NACRE}.  These rates are a crucial ingredient to BBN calculations.  
 
\subsection{Crucial BBN reaction rates}
As noted above, the usefulness of BBN \cite{Wagoner73, Yang84, Malaney93,Iocco09,Coc17,Cyburt16} as a cosmological constraint relies on the fact that the universe is in thermodynamic equilibrium at the relevant temperatures.  The 
only non-equilibrium process of relevance are the nuclear reactions themselves which must be explicitly evolved through the BBN epoch.  In all there are only
16 reactions of significance during BBN. \cite{Coc17,Nakamura17,Iocco09,Cyburt04,Cyburt16,Foley17} See however, the contribution by A. Coc and E. Vangioni to this volume \cite{Coc17} for suggestions on new reaction rates that may influence BBN.  Also, in Nakamura et al. \cite{Nakamura17} in this volume the broad range of reactions that can enter into the inhomogeneous big bang is summarized.  In order to be useful as a cosmological constraint one must know relevant nuclear reaction rates to very high precision ($\sim 1$\%).  Fortunately, unlike in stars, the energies at which these reactions occur in the early universe are directly accessible in laboratory
experiments. 
Although considerable progress has been made as discussed in the contributions in Refs.~[\refcite{Coc17,Nakamura17,Foley17}] to this volume, as well as in Refs.~[\refcite{Descouvemont04,Cyburt16}] in determining the relevant rates, much better rates are still needed for
the neutron life time \cite{Serebrov10,Mathews05}, the $^2$H$(p,\gamma)^3$He, $^2$H$(d,n)^3$He, $^3$He$(d,p)^3$He, 
$^3$He$(\alpha,\gamma)^7$Be, and $^7$Be$(n,\alpha)^4$He reactions.

\section{Key epochs of BBN}
The thermal history during BBN in reviewed in Refs~{\refcite{Coc17,Nakamura17}] in this volume.   We briefly here consider the three important time epochs of importance to nucleosynthesis:)
 \begin{itemize}
\item(a) At a time of $t\sim 10^{-2}$ s ($T \approx 10$ MeV) the weak interaction rates are faster than the universal
expansion rate.The energy density of the universe is dominated by relativistic particles $\gamma, \nu_i,  \bar \nu_i, e^+, e^-$, in thermal equilibrium.  Then from Eq.~(\ref{geff}) we have $g_{\rm eff} = 10.75$, and the neutron-to-proton
ratio is given by its equilibrium value, $n/p = \exp{(Ñ\Delta M/T)}$, where $\Delta M$  is the neutron-proton mass difference. Due to the high ambient temperature, the nuclear statistical equilibrium is shifted to a photodissociated nucleon gas and no nucleosynthesis occurs.
 
\item (b) As $t \rightarrow 1$ s ($T \approx 1$ MeV) the weak interaction rates can no longer maintain thermal equilibrium.  That is, until this time various weak reactions of electron neutrinos are in equilibrium, e.g.
$$n + \nu_e \leftrightarrows p + e^-$$
$$n + e^+  \leftrightarrows p + \bar \nu_e^-$$
$$n  \leftrightarrows p + e^- + \bar \nu_e $$
For the two-body weak reaction rates one can write:
\begin{equation}
\lambda_{np} = n_{\nu_e} \langle \sigma_{n+\nu_e} v \rangle + n_{e^+} \langle \sigma_{n+e^+} v \rangle ~~,
\end{equation}
where the number densities are given from Eq.~(\ref{ndens})
\begin{equation}
n_{\nu_e} = \frac{3 \zeta{3}}{4 \pi^2}T^3~~;~~ n_{e^+} = \frac{3 \zeta(3)}{2 \pi^2}T^3 ~~.
\end{equation}
From a quantum field theory derivation, one can obtain the averaged cross section:
\begin{equation}
\langle \sigma_{n+\nu_e} v \rangle = \frac{510 \pi^2}{ 3 \zeta(3) \tau_n Q^5} (12 T^2 + 6 QT + Q^2) ~~,
\label{nprate}
\end{equation}
and 
\begin{equation}
\langle \sigma_{n+e^+} v \rangle = \frac{255\pi^2}{ 3 \zeta(3) \tau_n Q^5} (12 T^2 + 6 QT + Q^2) ~~,
\end{equation}
where $Q$ is the neutron proton mass difference $m_n - m_p = 1.293$ MeV, and $\tau_n = 886.7$ s is the neutron lifetime.

Now, one expects the weak reaction to cease once the reaction rate per particle becomes slower than the expansion rate,
i.e $\lambda_{n p} < H$.
Adopting the leading term in Eq.~(\ref{nprate}), one can write for the most rapid of the weak reaction rates
\begin{equation}
\lambda_{n p} \approx 0.95 T^5 ~~{\rm s}^{-1} < H \approx 1.13 T^2 ~~{\rm s}^{-1} ~~,
\end{equation}
from which we deduce that the last of the weak reactions falls out of equilibrium for $T \approx 1.06$ MeV.

As long as the weak rates are in thermal equilibrium the ratio of neutrons to protons is simply related to the neutron-proton mass difference,
\begin{equation}
\frac{n}{p} = e^{-Q/T} 
\end{equation}
As the weak reactions fall quickly out of equilibrium then the $n/p$ ratio is Òfrozen inÓ at a value of $n/p \sim 1/6$. After this time the 
$n/p$ ratio is slightly  modified by free neutron decay to a value of about $1/7$ just before the onset of nuclear reactions.

 A crucial point to note here is that any increase in the ambient mass-energy density $\rho$  will lead to an increase in the expansion rate and a higher freeze-out temperature. This implies a higher
equilibrium value for the $n/p$ ratio when weak reactions freezeout.  This  ultimately leads to a higher $^4$He abundance when nucleosynthesis begins and almost all free neutrons are converted into $^4$He. 

\item (c) At $T \approx 0.5$ to $0.1$ MeV the $e^+-e^-$ pairs begin to annihilate heating the photon gas.  The neutrino gas, however, is unaffected by this time  because in the standard big bang all three neutrino flavors have already decoupled from the plasma by  $T \approx 1$ MeV. Because of this the photons are heated relative to the neutrinos by the simple fact that the entropy density $\propto g_{eff}T^{3}$ remains constant as the pairs annihilate and the effective number of degrees of freedom in Eq.~(\ref{geff}) diminishes.  Hence, the photons are hotter than neutrinos by a factor $(11/4)^{1/3}$ from this time onward. 

 To see how this temperature difference arises imagine and instantaneous annihilation of $e^+-e^-$ pairs.  Equating the entropy density just before and after annihilation  we have:
 \begin{equation}
 \biggl(2  + 2\frac{7}{8} + 3\frac{7}{8}\biggr)T_\nu^3  = 2 T_\gamma^3 + 3\biggl(\frac{7}{8}\biggr)T_\nu^3~~,
 \end{equation}
 where we label the temperature before as $T_\nu$ since up to this point the neutrino and photon temperatures are identical.  Afterward we distinguish between the neutrino temperature and photon temperatures.  The above includes  $g_\gamma = 2$ for the 2 helicity states of the photon and one must take into account the the neutrino is its own anti-particle, or that only left-handed neutrinos interact.  Then, removing the factor $3({7}/{8})T_\nu^3$ from both sides one obtains the ratio of the ratio of the photon to neutrino temperature:
 \begin{equation}
 \frac{T_\gamma}{T_\nu} = \biggl(\frac{11}{4}\biggr)^{1/3}~~.
 \end{equation}

\item (d) At $T \approx 0.1$  MeV ($t \approx 100$ s) the photodissociation rate of deuterium is slow enough for  the nuclear statistical equilibrium to shift to significant D production.   This occurs  along with the production of the other primordial isotopes $^3$He,
$^4$He and $^7$Li. However, the nuclear reaction rates diminish too rapidly to maintain nuclear statistical equilibrium. To calculate the nucleosynthesis yields, the time dependence of the temperature and density must be coupled to a network of nuclear reaction rates \cite{Wagoner73, Cyburt16, Foley17} as noted above. The coupled differential equations Eq.~(\ref{nucrates})  must then be integrated until the density and temperature fall to the point where all nuclear reactions cease.

In essence, however, essentially all free neutrons will be converted into $^4$He nuclei, with a small fraction converted to other elements.  Hence, it is straightforward to estimate the $^4$He mass fraction.  Since each neutron absorbed into $^4$He will also extract a proton, and $m_p \approx m_n$, then the helium mass fraction will be:
\begin{equation}
Y_p = \frac{2 n}{n + p} = \frac{2}{1 + p/n} ~~.
\end{equation}
Since by $t \approx 100$ seconds the n/p ratio has diminished from 1/6 to 1/7 by neutron decay, then the helium mass fraction becomes $Y_p \approx 0.25$.  This is very close to the observed primordial abundance os $Y_p = 0.2449 \pm 0.0040$ as noted in the next section.
 \end{itemize}

\section{Light Element Abundances}
One of the powers of standard-homogeneous BBN is that once the reaction rates are known, all of the light element abundances are determined
in terms of the single parameter $\eta$ (or $\eta_{10} $ defined as  the baryon-to-photon ratio in units of $10^{10}$). The crucial test of the
standard BBN is, therefore, whether the independently determined value of  $\eta_{10}$ from fits to the CMB reproduces all of the observed primordial
abundances. There are reviews of this in this volume \cite{Coc17,Nakamura17} and elsewhere \cite{Iocco09,Cyburt16}, and also new constraints on the primordial
helium abundance \cite{Aver10,Aver15}. The best abundance constraints for this review are then adopted from Ref.~[\refcite{Cyburt16,Coc17,Nakamura17}] as follows:

\subsection{Deuterium}
Deuterium is best measured in the spectra of narrow-line Lyman-$\alpha$  absorption systems in the foreground of high
redshift QSOs. Unfortunately, only about a dozen of such systems have been found \cite{Iocco09,Cyburt16}. Taken altogether they exhibit an
unexpectedly large dispersion. This suggests that there could be unaccounted systematic errors. This enhanced
error can be approximately accounted for by constructing the weighted mean and standard deviation directly from
the data points. Based upon this, a conservative range for the primordial deuterium abundance of
D/H $= (2.87^{+0.22}_{-0.19})\times 10^{-5}$ from \cite{Coc17}. This implies a $2 \sigma$ (95\% C.L.) concordance region of:
$2.49 \times 10^{-5} < D/H < 3.3 \times 10^{-5}$
We note, however, that if one restricts the data to the six well resolved systems for which there are multiple 
Lyman-$\alpha$ lines\cite{Coc17,Pettini12,Cooke14,Cooke16,Cyburt16}, one slightly lowers the deuterium constraint to the presently adopted value \cite{Cyburt16} of 
\begin{equation}
{\rm D/H} = 2.53 \pm 0.04 \times 10^{-5} .
\end{equation}
\subsection{$^3$He}
The abundance of $^3$He is best measured\cite{Bania02} in Galactic HII regions by the 8.665 GHz hyperfine transition of $^3$He$^+$. A
plateau with a relatively large dispersion with respect to metallicity has been found at a level of $^3$He/H $=(1.90 \pm 0.6) 
\times 10^{-5}$. It is not yet understood, however, whether $^3$He has increased or decreased through the course of stellar and
galactic evolution \cite{Chiappini02,Vangioni-Flam03}. Whatever the case, however, the lack of observational evidence for the predicted galactic
abundance gradient \cite{Romano03} supports the notion that the cosmic average $^3$He abundance has not diminished from that
produced in BBN by more than a factor of 2 due to processing in stars. This is contrary to the case of deuterium for
which the observations and theoretical predictions are consistent with a net decrease since the time of the BBN epoch.
Moreover, there are results \cite{Eggleton06} from 3D modeling of the region above the core convective zone for intermediate-mass
giants which suggest that in net, $^3$He is neither produced nor destroyed during stellar burning. Fortunately, one can
avoid the ambiguity in galactic $^3$He production by making use of the fact that the sum of (D +$^3$He)/H is largely
unaffected by stellar processing. This leads to a best estimate \cite{Iocco09,Coc17} of $^3$He/H $= (0.7 \pm 0.5) \times 10^{-5}$ which implies a
reasonable 2$\sigma$ upper limit of
$^3$He/H $ < 1.7 \times 10^{-5}$ 
and a lower limit of zero.

\paragraph{$^4$He}
The primordial $^4$He abundance, Yp is best determined from HII regions in metal poor irregular galaxies extrapolated
to zero metallicity. A primordial helium abundance of $Y_p = 0.247 \pm 0.002 {\rm~stat} \pm 0.004 {\rm~syst}$ was deduced in Ref. \cite{Iocco09}
based upon an analysis \cite{Piembert07} that included new observations and photoionization models for metal-poor extragalactic
HII regions, along with new atomic physics computations of the recombination coefficients for HeI and the collisional
excitation of the HI Balmer lines. However, the extraction of the final helium abundance has been fraught with uncertainties
due to correlations among errors in the neutral hydrogen determination and the inferred helium abundance.
In \cite{Aver10,Aver15} it was demonstrated that updated emissivities and the neutral hydrogen corrections generally increase the inferred
abundance, while the correlated uncertainties increase the uncertainty in the final extracted helium abundance.
Therefore, we adopt the value and uncertainty from \cite{Aver15} of 
\begin{equation}
{\rm Y_p }= 0.2449 \pm 0.0040 , 
\end{equation}
which is in general agreement with
the predicted value from standard BBN when $\eta_{10}$ is fixed  from the {\it Planck}  analysis \cite{PlanckXIII}. 

\paragraph{$^7$Li}
The primordial abundance of $^7$Li is best determined from old metal-poor halo stars at temperatures corresponding
to the Spite plateau (see Refs.~[\refcite{Iocco09,Coc17,Cyburt16}] and Refs. therein). There is, however, an uncertainty in this determination due to the
fact that the surface lithium in these stars may have experienced gradual depletion due to mixing with the higher temperature
stellar interiors over the stellar lifetime. On the other hand, there are limits on the amount of such
depletion that could occur since most lithium destruction mechanisms would imply a larger dispersion in abundances
determined from stars of different masses, rotation rates, magnetic fields, etc., than that currently observed. In view
of this uncertainty a reasonable upper limit on the $^7$Li abundance has been taken \cite{Iocco09} to be $6.15 \times 10^{-10}$ which is based
upon allowing for a possible depletion of up to a factor of $\sim$5 down to the present observationally determined value.  For our purposes we adopt  
\begin{equation}
 ^7{\rm Li/H} = 1.6 \pm 0.3 \times 10^{-10} ~~,
 \end{equation}
 based upon an average of halo stars with [Fe/H]$\ge -3$   from Ref.~[\refcite{Cyburt16}].

\begin{figure}%[ht]
\centering
\includegraphics[scale=.7]{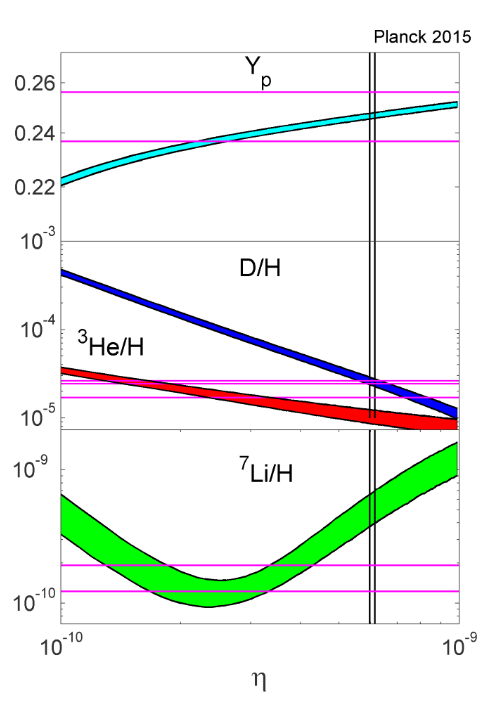}
\caption{BBN abundances as a function of the  baryon to photon ratio.  Shaded bands  correspond to the $2 \sigma$ (95\% C.L.) uncertainties deduced in 
[25].  
Horizontal lines show range of the uncertainties in the primordial abundances inferred in 
[12]. 
Vertical lines indicate the value of $\eta$ deduced in the {\it Planck} analysis [48].%\cite{PlanckXX}
}
\label{fig:abund}
\end{figure}

\section{Comparison of BBN with Observed abundances}

Ultimately, the value of BBN is in the detailed comparison between the  primordial abundances inferred from observation and the predictions based upon the BBN calculation.
Several papers in this volume\cite{Coc17,Nakamura17,Foley17} address this comparison.  Figure \ref{fig:abund} from Ref.~[\refcite{Foley17}] summarizes the current situation.  This is the famous "Schramm plot\cite{Yang84}" showing primordial abundances as a function of the baryon to photon ratio $\eta$.  Shaded bands show the uncertainty in the BBN abundances\cite{Foley17} while the horizontal lines show the uncertainties in the observationally inferred primordial abundances. Vertical lines show the constraints $\eta = (6.10 \pm0.04)\times 10^{-10}$ deduced from the {\it Planck} analysis \cite{PlanckXX} of the cosmic microwave background.  One of the amazing triumphs of BBN is the overall agreement between the predictions of BBN and the observed primordial abundances over nine orders of magnitude in abundances.  In particular the observed D/H abundance agrees almost perfectly with the BBN prediction in within the CMB inferred value of $\eta$.   Similar agreement exists for $Y_p$ and the $^3$He/H ratio, although the observational uncertainty is much larger.  

  There is, however, one glaring problem that remains.  The calculated and observed $^7$Li/H ratio differ by about a factor of 3.  This is known as "the lithium problem."  Several of the papers in this issue address this problem.\cite{Coc17,Nakamura17,Kusakabe17,Sato17,Yamazaki17}  At present it is not yet known if this discrepancy derives from a destruction of Lithium on the old stars used to deduce the primordial Lithium abundance, or if it requires exotic new physics in the early universe\cite{Coc17,Nakamura17,Kusakabe17,Sato17,Yamazaki17}, or even a modification of the particle statistics in BBN itself \cite{He17}.  For the remainder of this introduction we will adopt the premise that this disagreement may indicate new physics in the early universe.  It is important then to address what are the key questions in modern cosmology and how might BBN address these questions.

\section{What are the  Questions?}
 
The important questions regarding the big bang are something like the following. Some of these 
intriguing questions  will be addressed  below:
\begin{enumerate}
\item How did the universe begin?
\item Why are there 3 large spatial dimensions?
\item  What drives inflation?
\item  Are there observable effects from: supersymmetric particles, string excitations, etc?
\item  Is there evidence for large extra dimensions?
\item  How does the universe reheat?
\item  How and when was the net baryon number generated?
\item  When and how was the dark matter generated?
\item  When and how was the dark energy created?
\item  Are there observable effects from the Electroweak or QCD transition?
\item  Have the fundamental constants varied with time?
\item  Is there a primordial magnetic field?
\end{enumerate}
\subsection{What drives Inflation?}
The simplest explanation for the fact that the universe is so nearly 
at today ($\Omega_{0} = 1.000 \pm 0.005$  Ref.~[\refcite{PlanckXIII}]) and the near
isotropy of CMB is to conclude  that the universe has gone through an epoch  rapid inflation.\cite{Starobinsky,Guth, Linde}
 The simplest  view 
is that some vacuum energy $V(\phi)$ drives inflation due to the existence of a self-interacting scalar field $\phi$. That is, the
energy density of the cosmic  fluid in the early universe includes a dominant contribution from the evolution of a scalar field.
The mass energy density for a scalar field can be deduced from the Klein-Gordon Equation
\begin{equation}
\rho(\phi) = \frac{\dot \phi^2}{2} + \nabla^2 \phi   + V(\phi)~~.
\end{equation}
The inflaton  field $\phi$ itself evolves according to a damped harmonic-oscillator-like equation of motion:
\begin{equation}
\ddot \phi + 3 H \dot \phi - \nabla^2 \phi + dV/d\phi = 0~~.
\label{ddotphi}
\end{equation}
As the
universe expands, $H$ is large and the $\dot \phi$ term dominates as a kind of friction term. The universe is then temporarily   trapped in a slowly varying $V(\phi)$ dominated regime so that 
the scale factor grows exponentially.  This is known as the slow roll approximation for which in the absence of spatial fluctuations Eq.~(\ref{ddotphi}) becomes
\begin{equation}
\dot \phi =  \frac{dV/d\phi}{ 3 H}~~.
\end{equation}

The biggest unknown quantity in this paradigm is the form of $V(\phi)$. The simplest form $V(\phi) = (m/2) \phi^2$ may be motivated
by the Kahler potential in string theory, however, almost any form for the potential works well to describe the big bang. 
 Recently, the determination\cite{PlanckXX} ratio of the tensor to scalar contributions to the CMB power spectrum have ruled out many of the simplest forms of the 
 inflation generating potential.  Indeed, the only monomial type potentials that are marginally consistent with the data are those motivated by string theory
 such as the lowest order axion monodromy potentials\cite{McAllister10, Silverstein08} with $V(\phi) \propto \phi^{2/3}$ or $\propto \phi$, natural inflation, \cite{Freese90, Adams93, Freese93}
 or the $R^2$ inflation.\cite{Starobinsky80}
 
 For the most part, BBN is unaffected by the inflation-driving potential except in the special case of quintessential inflation.\cite{Peebles}  In Ref.~[\refcite{Yahiro02}] we looked at 
 this intriguing attempt to reduce the inflation potential problem, the baryogenesis question, and the dark energy mystery into a
single paradigm which involves non-minimal coupling between matter and gravity as the universe makes a transition
from an inflation driving potential to a dark-energy producing quintessence. Big bang nucleosynthesis significantly
constrains this paradigm as the non-minimal couplings lead to an excess energy density in gravity waves which alter
the results of BBN.

\subsection{Is there Evidence for Large Extra Dimensions?}
In one form of the low-energy limit to M-theory,\cite{Mtheory1,Mtheory2} the universe can be represented by two 10-dimensional manifolds separated by a large extra dimension. It is possible
that the extra dimension could manifest itself on the dynamics of the universe and BBN \cite{Ichiki02,Sasankan17}. For example, in a
Randall-Sundrum II \cite{Randall99} brane-world cosmology, the cosmic expansion for a 3-space embedded in a higher dimensional
space can be written \cite{Ichiki02,Sasakan17} as
\begin{equation}
\biggl( \frac{ \dot a}{a}\biggr)^2 = \frac{8}{3} \pi G \rho -\frac {k}{a^2} + \frac{\Lambda}{3} 
+ \frac{\kappa_5^4}{36} \rho^2 + \frac{\mu}{a^4}~~,
\label{Friedman2}
\end{equation}
where the four-dimensional gravitational constant $G_N$ is related to $\kappa_5$ the five-dimensional gravitational constant,
i.e. $G_N {\kappa_5^4 \lambda}/{48 \pi}$, with $\lambda$ the intrinsic tension of the brane. The fourth term arises from the imposition of a junction
condition for the scale factor on the surface of the brane, and is not likely to be significant. The fifth  term, however,
scales just like radiation with a constant $\mu$ and is called the {\it dark radiation}. Its magnitude and sign derives from the
projection of curvature in higher dimensions onto four-dimensional space-time. Because this dark radiation scales as $a^{-4}$
it can affect both BBN and the CMB. It can significantly alter\cite{Ichiki02,Sasakan17} the fit to BBN abundances and the CMB, and hence can be constrained.
In \cite{Sasakan17} it is shown that the newest light element abundance constraints and reaction rates significantly limits the possibility of brane-world dark radiation.

\subsection{When and how was the dark energy created?}

There are a variety of models for the dark energy besides that of a simple cosmological constant. Dark energy
can also be attributed to a vacuum energy in the form of a "quintessence" scalar field which must be slowly evolving
along an effective potential. A quintessence or k-essence field is of interest as it can be constrained by both BBN
and the CMB \cite{Ichiki02}. 

However, the simple coincidence that both of dark matter and dark energy currently contribute
comparable mass energy toward the closure of the universe begs the question as to whether they could be different
manifestations of the same physical phenomenon. 
If the dark matter is flowing form a higher dimension,\cite{Umezu} or the dark matter generates a cosmic bulk viscosity.\cite{WilsonBV,Mathews08}
Another possibility is that an  inhomogeneous distribution of dark matter might produce relativistic corrections
to the Friedmann equation (\ref{Friedmann}) that lead to a dark-energy like term.  It seems likely, however, that such corrections are small.\cite{Zhao11,Mathews08}

\subsection{Is there evidence of supersymmetric matter in the early universe?}
Although supersymmetry is a well motivated mathematical symmetry, the first runs at the Large Hadron Collider have not found evidence for its existence. 
Nevertheless, there are reasons to remain optimistic about its validity in Nature.\cite{Ellis15}  
For example, the lightest stable supersymmetric particle is still a very good candidate for the cold dark matter.  If that is the case,\cite{Kusakabe17,Sato17}
then many other unstable supersymmetric particles would have been generated along with the dark matter in the very early universe.
This issue is taken up in the papers of this volume by Kusakabe et al./cite{Kusakabe17}, Nakamura et al.\cite{Nakamura17}, and Yamazaki et al.\cite{Yamazaki17} In particular, the next to lightest supersymmetric particle could have a lifetime long enough to be present and/or decay during BBN.
 If that is the case then the non-thermal decay of SUSY particles or the presence of a long-lived negatively charged  particle (e.g. the supersymmetric
partner of the electron or tau) 
 might explain the over-production of $^7$Li during BBN.\cite{Kusakabe14,Kusakabe17,Sato17,Yamazaki17}  
 
 At one time such models were motivated by an inferred high primordial $^6$Li abundance in metal poor stars.\cite{Asplund06}  However, that has  now 
 been shown \cite{Lind13} to
 be an artifact of 3D effects in stellar atmospheres and non-local thermodynamic equilibrium (NLTE) in models to fit the lithium absorption line profile.
 Nevertheless, the observed upper limit to the $^6$Li ($\sim 2$\% of $^7$Li abundance) still places important constraints on 
 supersymmetric models and there remains evidence \cite{Howk12} for primordial $^6$Li in the SMC.  
   
 As noted in Section 3, the  $^7$Li is as much as a factor of 3 below the
BBN expectation.  This might be a  manifestation of the existence of new unstable
particles which decay during and/or after the big bang \cite{Pospelov,Kusakabe07,Kusakabe10,Kusakabe11,Kusakabe14}.
In particular, a number of  papers \cite{Pospelov,Cyburt06,Kaplinghat06,Kohri06,Bird07,Kusakabe07,Kusakabe10,Kusakabe11,Kusakabe14} including three in this volume \cite{Kusakabe17,Sato17,Yamazaki17} have 
considered heavy negatively charged decaying $X^-$ particles that modify BBN. The heavy $X^-$ particles bind to the nuclei produced in BBN. The
massive $X^-$ particles reduce the reaction Coulomb barriers and enhance the thermonuclear reaction rates.  These effects extend
the duration of BBN to lower temperatures. This can lead to an enhancement of the $^6$Li abundance \cite{Kusakabe07} (for
example by the $^4$He$_X(d,X^-)^6$Li reaction, while depleting $^7$Li.

\subsection{Is there evidence of a QCD phase transition?}
Many papers \cite{Nakamura17,Malaney93, Lara06} have considered the possibility that BBN could constrain the details of a first order QCD transition
in the early universe. However, results from lattice gauge theory calculations seem to rule out \cite{LGT}
the possibility of a first-order transition at low baryon density.  Moreover, the CMB  limits on the baryon-to-photon ratio imply such tight constraints on
the the allowed inhomogeneous big bang parameters and nucleosynthesis\cite{Lara06}, that it is probably not possible to have
a significant effect on BBN from a first order primordial QCD transition.  Nevertheless, the possibility of significant baryon inhomogeneities during BBN arising from other phase transitions
remains a viable possibility as discussed in the contribution by Nakamura et al.\cite{Nakamura17} to this volume.

\subsection{Have the fundamental constants varied with time?}
A time dependence of fundamental constants in an expanding universe often arises\cite{Uzan03,Flambaum08} in  theories
that attempt to unify gravity and other interactions. Such grand unified theories imply correlations among variations
in all of the fundamental constants \cite{Flambaum08}.  BBN is sensitive, for example, to  variations in the fine structure constant\cite{Dmitriev07,Iocco09}.
However, it can be argued \cite{Flambaum07} that BBN is much most sensitive to variations
in the dimensionless quantity $X_q \equiv m_q/\Lambda_{QCD}$, where $m_q$ is the average quark mass, and $\Lambda_{QCD}$ is the scale of quantum chromo dynamics.
 Hence, $X_q$ may be the
best parameter with which to search for evidence of time variation of fundamental constants in the early universe.

In one study \cite{Berengut10} it was deduced  that an increase in the average quark mass by an amount
$\delta m_q/m_q =0.016 \pm 0.005$ provides a better agreement between observed light-element primordial abundances than
those predicted by the standard big bang. A similar conclusion was reached in \cite{Dent07}. Because of the importance of
such evidence for a changing quark mass, it is important to carefully reexamine all aspects of the physics which has
been used to place constraints upon this parameter from BBN. 

In addition to the interest in evidence for a variation of the physical constants in the early universe, it has also been
suggested \cite{Flambaum07,Berengut10} that such variations may provide insight into a fundamental problem in BBN. As noted above there is an apparent
discrepancy between the observed primordial abundance of $^7$Li and that inferred from BBN when the limits on the
baryon-to-photon ratio $\eta$ from the WMAP 9yr or Planck  analyses are adopted.

In  Ref.~[\refcite{Cheoun11}] an independent evaluation was made of the effects on BBN from a variation in
the parameter $\delta X_q/X_q$. This constraint is very dependent upon the detailed analyses \cite{Cyburt16,Foley17} of the uncertainties in the
observed light element abundance constraints. An independent evaluation\cite{Cheoun11} of the resonant $^3$He$(d, p)^4$He
reaction rate based upon both the forward and reverse reaction dependence on $\delta m_q/m_q$.   Although the
uncertainty in the results increases due to variations in the resonance parameters, the newer abundance constraints
narrow the range of possible variations in the quark mass from BBN. in Ref.~[\refcite{Cheoun11}] it was deduced that this latter constraint dominates the 
results are  no variation in the averaged quark mass.

\subsection{Is there a primordial magnetic field?}

The existence of a primordial magnetic field (PMF) is discussed in the contribution by Yamazaki et al.\cite{Yamazaki17} in this volume.   
A present-day cosmic magnetic field of $\sim$1 nG whose field lines collapse as structure forms is one
possible explanation for the magnetic fields observed in galactic clusters. Such a PMF, however, could have influenced
a variety of phenomena in the early universe \cite{Grasso01} such as the cosmic microwave background (CMB), (e.g. \cite{Yamazaki12,Yamazaki17} and refs. therein). In
a series of papers\cite{Yamazaki06,Yamazaki08,Yamazaki10a,Yamazaki10b,Yamazaki12,Yamazaki17}  the  correlations 
between the calculated CMB power spectrum (influenced by a PMF) and the
primary curvature perturbations were analyzed.  It was found  found that the PMF affects the CMB on both small and large angular scales in
the TT and TE modes. At that time the introduction of a PMF led to a better fit to the CMB power spectrum for the higher
multipoles, and the fit to the lowest multipoles could be used to constrain the correlation of the PMF with the density
fluctuations. The best constraints on the PMF determined determined in that  analysis was
$\vert B\vert < 2.10$ nG (68\%CL) ; $< 2.98$ nG (95\%CL)
on a present scale of 1 Mpc, and
$n_B<  1.19$ (68\%CL) ; $< 0.25$ (95\%CL)
It was  found that the BB mode is dominated by the vector mode of the PMF for higher multipoles. It was also shown
that by fitting the complete power spectrum one can break the degeneracy between the PMF amplitude and its power
spectral index.

Of particular relevance to this review, however, is that a primordial magnetic field can lead to fluctuations in the metric.
The implied background in the induced gravity waves can then be used\cite{Yamazaki12} to constrain the epoch at which the
PMF was created \cite{Caprini01,Caprini09}.
The balance between the expansion rate of the universe and various  particle reaction rates has important effects on the nucleosynthesis of  light elements in the big-bang.
Moreover, since the energy density of the gravity wave background, $\rho_{\mathrm{GW}}$,
contributes to the total energy density of the universe,
the expansion rate is affected by the GWB.
Therefore, one can indirectly constrain the energy density  $\rho_{\mathrm{GW}}$ from the  light element primordial abundances inferred from observations  of deuterium (D), $^3$He, $^4$He, and $^7$Li.  
%The BBN constraint on the gravity wave background due to a primordial magnetic field becomes :
%\begin{eqnarray}
%\int^\infty_0
%	d\log{(\nu)}
%	h^2_0
%	\Omega_\mathrm{GWB}(\nu)
%&&\le
%4.7\times 10^{-6} \mathrm{BBN,~D~and~^4He},\\
%\label{eq:gw_Nn}
%\end{eqnarray}
Based upon the constraints on the parameters of the magnetic field it was concluded in Ref.~[\refcite{Yamazaki12}]
that BBN most favors  a magnetic field formed after the BBN epoch, although  no earlier epoch is yet ruled out. 

\section{Conclusion}
In this short review it is hoped that the reader has gained adequate exposure basic concepts and modern applications of big bang nucleosynthesis to topics in modern cosmology.
The reader should now be prepared to review the more detailed applications to be found in the subsequent contributions to  this volume.

\section*{Acknowledgments}
Work at the University of Notre Dame was
 supported by the U.S. Department of Energy under Nuclear Theory Grant
 DE-FG02-95-ER40934.  This work has been supported in part by Grants-in-Aid for Scientific
Research (24340060, 20244035), and on Innovative Areas (20105004) 
of the Ministry of Education, Culture, Sports, Science and Technology of Japan, and in part by Grants-in-Aid for Scientific Research of JSPS (24340060), and Scientific Research on Innovative Areas of MEXT (20105004).

\end{document}